Estimating financial risk measures for futures positions: a non-parametric approach

By

John Cotter and Kevin Dowd[*]

Abstract

This paper presents non-parametric estimates of spectral risk measures applied to long and short positions in 5 prominent equity futures contracts. It also compares these to estimates of two popular alternative measures, the Value-at-Risk (VaR) and Expected Shortfall (ES). The spectral risk measures are conditioned on the coefficient of absolute risk aversion, and the latter two are conditioned on the confidence level. Our findings indicate that all risk measures increase dramatically and their estimators deteriorate in precision when their respective conditioning parameter increases. Results also suggest that estimates of spectral risk measures and their precision levels are of comparable orders of magnitude as those of more conventional risk measures.

Running head: financial risk measures for futures positions

December 23, 2006

[*] John Cotter is at the Centre for Financial Markets, Smurfit School of Business, University College Dublin, Carysfort Avenue, Blackrock, Co. Dublin, Ireland; email: john.cotter@ucd.ie. Kevin Dowd (corresponding author) is at the Centre for Risk and Insurance Studies, Nottingham University Business School, Jubilee Campus, Nottingham NG8 1BB, UK; email: Kevin.Dowd@nottingham.ac.uk. Cotter's contribution to the study has been supported by a University College Dublin Faculty of Commerce research grant. Dowd's contribution was supported by an Economic and Social Research Council research fellowship on 'Risk measurement in financial institutions' (RES-000-27-0014).

## 1. INTRODUCTION

One of the more interesting developments in financial risk management in the last few years is the theory of spectral risk measures (see Acerbi (2002, 2004)). Spectral risk measures (SRMs) satisfy the properties of coherence, and therefore have all the attractions of the coherent risk measures that appeared a little while earlier (see Artzner *et alia* (1999)). However, unlike more conventional risk measures such as the VaR or the Expected Shortfall (ES), spectral measures also take account of the user's risk aversion. Indeed, SRMs are the only risk measures that are both coherent and take explicit account of the degree of user risk-aversion. They are also an important class of risk measures that have many possible applications, not least in situations where risk factors are very non-normal and conventional portfolio theory leads to unreliable measures of financial risk. However, to date there are very few estimates of SRMs available, and their empirical properties are not well understood.[1]

Each of these three risk measures depends on a key conditioning parameter. In the case of the VaR and the ES, the conditioning parameter is the confidence level; and in the case of the spectral risk measures considered here, the conditioning parameter is the coefficient of absolute risk aversion (ARA).[2] The fact that the VaR and the ES have the same conditioning parameter makes it very easy to compare them; however, it is also possible to make some comparisons between these two risk measures and spectral ones and, in particular, to compare how these measures change in the face of varying conditioning parameters.

These risk measures also have promising potential applications to futures markets. The most obvious is that they can be used to determine the margin requirements on futures positions.[3] For example, estimates of these risk measures applied to unconditional price (or return) distributions provide a natural basis on which futures clearinghouses might determine the initial margins on futures positions, and one can argue that these risk measures would provide a superior basis for setting initial margins than the SPAN systems that are currently used for this purpose.[4] Similarly, estimates of these risk measures applied to conditional price distributions (e.g., using GARCH modelling approaches) would provide a natural basis on which clearinghouses might determine the corresponding maintenance margins. However, in this paper we restrict ourselves to the simpler unconditional problem, and leave the more difficult conditional modelling problem to a later paper.

More specifically, this paper presents non-parametric estimates of these various risk measures applied to long and short positions in 5 of the most prominent equity index futures contracts – the S&P500, the FTSE100, the DAX, the Hang Seng, and the Nikkei225. As these risk measures each have a conditioning parameter, it is more important to examine how they behave over a range of possible values that these parameters might take. Moreover, since the usefulness of any risk measure estimate also depends on its precision, the paper also reports results for the precision of these estimators. We note here that non-parametric estimation methods have the advantages that they do not require us to make potentially questionable assumptions about the distributions governing futures returns, and they are more straightforward to apply.

Our empirical findings suggest four main conclusions. First, over the ranges of conditioning parameters considered here, estimates of ES tend to be a little larger than those of the VaR, and estimates of SRMs are not massively different from estimates of either. Second, estimated risk measures rise quite sharply as the conditioning parameter rises. The fact that the risk measure rises with its conditioning parameter is to be expected, but the sensitivity of the risk measure to its parameter is an empirical issue, and this sensitivity is quite pronounced. Thirdly, we find that there are no great differences in estimates of the precision of different risk measure estimators: ES estimators are perhaps a little less precise than comparable VaR estimators, and SRM estimators have much the same precision as VaR ones. Finally, we also find that the precision of these estimators falls as the conditioning parameter gets larger: estimators of VaR or ES become less precise as the confidence level rises, and estimators of SRMs become less precise as the risk aversion increases. As we shall see, these results can be explained in terms of decreasing effective sample size.

The remainder of the paper is organized as follows. Section 2 examines the risk measures estimated in this paper. Section 3 provides a description of the data and of the main empirical features of the five futures contracts chosen for analysis, and section 4 discusses estimation methods. Sections 5, 6 and 7 present results for the VaR, the ES and the spectral measures in turn. Conclusions are offered in section 8.

**2. MEASURES OF RISK**

Consider the realized loss on a futures position (which is positive for an actual loss, and negative for a profit). If the confidence level is $\alpha$, the VaR at this confidence level is:

$$VaR_\alpha = q_\alpha \tag{1}$$

where $q_\alpha$ is the $\alpha$-quantile of the loss distribution. The VaR is the most widely known financial risk measure, but has been widely criticized in recent years. The most serious criticisms are that it does not satisfy the property of subadditivity and therefore lacks coherence (Artzner *et al.*, 1999), and that it can give unreliable assessment of risk exposure because it takes no account of losses beyond the tail or VaR threshold.

The ES risk measure avoids both these problems: the ES is both subadditive and coherent, and it takes account of the sizes of losses in the tail beyond the VaR itself. The ES measure is defined as the average of the worst $1-\alpha$ of losses. In the case of a continuous loss distribution, the ES is given by:

$$ES_\alpha = \frac{1}{1-\alpha}\int_\alpha^1 q_p\, dp \qquad (2)$$

The ES gives equal weight to each of the worst $1-\alpha$ of losses, and gives no weight to any other observations. However, like the VaR, the ES is dependent on an arbitrarily chosen confidence level and there is little a priori to tell us what value this should take.

In addition, neither of these risk measures takes any explicit account of a user's degree of risk aversion.[5] In fact, it turns out that the choice of the VaR implies that the user has *negative risk aversion*, and the choice of the ES implies that the user is *risk-neutral* (see, e.g., Grootveld and Hallerbach (2004)). The negative risk aversion of the VaR is illustrated by the fact that the user places no weight on losses exceeding VaR, and the risk-neutrality of the ES is illustrated by the fact that the user places equal weight on losses exceeding the VaR. Thus, neither of these risk measures is consistent with the user being risk-averse.

To find a risk measure that is consistent with the user's risk aversion, we now examine the spectral risk measures proposed by Acerbi (2002). Consider a risk measure $M_\phi$ defined by:

$$M_\phi = \int_0^1 q_p \phi(p) dp \qquad (3)$$

where $q_p$ is the $p$ loss quantile, $\phi(p)$ is a weighting function defined over $p$, and $p$ is a range of cumulative probabilities $p \in [0,1]$. Following Acerbi (2004), the risk measure $M_\phi$ is coherent if and only if $\phi(p)$ satisfies the following properties:

- $\phi(p) \geq 0$: weights are always non-negative.

- $\int_0^1 \phi(p) dp = 1$: weights sum to one.

- $\phi'(p) \geq 0$: higher losses have weights that are higher than or equal to those of smaller losses.

We now need to specify a suitable weighting (or risk-aversion) function, and one plausible choice is an exponential risk-aversion function:

$$\phi(p) = \frac{ke^{-k(1-p)}}{1-e^{-k}} \qquad (4)$$

where $k>0$ is the coefficient of absolute risk aversion. This weighting/risk-aversion attaches higher weights for larger losses at higher cumulative probability levels; moreover, the weights rise more rapidly as the user becomes more risk-averse, and both these features are exhibited in Figure 1 for a spectrum of absolute risk aversion coefficients.

**Insert Figure 1 here**

## 3. PRELIMINARY DATA ANALYSIS

Our data consist of daily log difference returns for five of the most liquid futures: S&P500, FTSE100, DAX, Hang Seng and Nikkei225 indexes. The data cover the period between January 1 1991 and December 31 2003. The data were obtained from Datastream with the contracts trading on the CME (in the cases of the S&P500 and Nikkei 225), LIFFE (in the case of the FTSE100), EUREX (in the case of the DAX) and HKSE (in the case of the Hang Seng). These data refer to futures contracts with a rollover from an expiring contract to the next one occuring at the start of each new contract cycle. As Datastream deals with bank holidays by padding the dataset and taking the bank holiday's end-of-day price to be the previous trading day's end-of-day price, this means that we have the same number of daily returns (i.e., 3392) for all contracts.

Summary statistics for these are given in Table 1, which gives the first four moments and max/min statistics for each of these return series. Average returns are generally positive except for the Nikkei 225 futures, and daily volatility ranges between approximately 1% (for the S&P500) and 2% (for the Hang Seng). Excess skewness are of differing signs (negative for the American and European, positive for the Asian) which suggests that there may be differences between the risks of long positions and corresponding short positions in these contracts. All the contracts also exhibit excess kurtosis, but the amount of excess kurtosis varies considerably (from 1.78 for the Nikkei to nearly 11 for the HangSeng). There are also large differences in the range between sample minimum and maximum values. Overall, these summary results suggest that the distributions of equity futures returns are non-normal, and vary somewhat from one contract to another.

**Insert Table 1 here**

## 4. NON-PARAMETRIC ESTIMATION

We estimate our risk measures using a vanilla non-parametric bootstrap.[6] A non-parametric bootstrap is more robust and more flexible than a parametric one, because it does not depend on questionable parametric assumptions that we are not in a strong position to make.[7] A non-parametric bootstrap also has the advantage of making it easy for us to estimate the precision metrics (e.g., standard errors or confidence intervals) for each of our three risk measures. As explained in standard references, the bootstrap enables us to estimate standard errors or confidence intervals for any parameter that we can estimate from sample data (e.g., Efron and Tibshirani, 1993; or Davison and Hinckley, 1997). Such precision indicators are more difficult to estimate when using parametric estimation approaches.[8]

For the purposes of our empirical analysis, we estimate VaR and ES using confidence levels spanning the range 90% to 99%, and we estimate SRMs using coefficients of absolute risk aversion spanning the range from 5 to 80 as the associated risk measures exhibit similar magnitudes for these values.

**5. RESULTS FOR VAR**

Table 2 presents some bootstrap-average results for estimated VaRs in long and short positions over various common confidence levels. Section (a) of the Table shows sample estimates of VaR. For instance, the first item on the left shows that the daily VaRs on a long position in the S&P500 futures at a 90% confidence level is 1.19%. The results in this section also show that the VaRs increases considerably as the confidence level gets bigger. This is illustrated by the fact that the average VaR (in the rightmost column) increases from 1.61% for the 90% confidence level to 3.82% for the 99% confidence level. The VaRs also vary considerably with the contract: for example, the VaRs are lowest for the S&P and FTSE contracts indicating lower inherent risk in US and UK markets, and highest for the Hang Seng indicating higher risk for the Hong Kong market.

**Insert Table 2 here**

As well as being concerned with the point estimates of our risk measures, we also need to examine their precision. We use three metrics to assess the precision of each estimated risk measure, namely, the standard error, the coefficient of variation, and the 90% confidence interval.

The standard errors for futures VaRs are given in section (b) of Table 2. Similar to the VaRs, the standard errors are highest for the Hang Seng futures, regardless of trading position and confidence level. The magnitudes of standard errors follow the same pattern as the VaRs, and increase markedly with the confidence level. On average, the standard error for a long position increases from 0.05% for VaR at the 90% confidence level to 0.17% for VaR at the 99% confidence level – an increase of over 200%. These results indicate that the precision of our VaR estimates falls markedly as the confidence level rises.

A drawback with the standard error is that it is an *absolute* measure of precision and makes no allowance for the size of the standard error relative to the size of the estimated risk measure. Since VaR rises with the confidence level, it is therefore arguable that a mere comparison of standard errors can give a misleading impression of the precision of our VaR estimates. If we wish to assess the *relative* precision of our estimates, a better indicator is the coefficient of variation, which is the ratio of a point estimate to the corresponding standard error. The coefficient of variation therefore gives us an estimate of the precision of our estimates that takes account of the size of the point estimate itself. Some estimates of the coefficient of variation of the VaR are given in section (c) of Table 2. The coefficients of variation falls for increasing confidence levels, which again suggests that precision falls as the confidence level rises; however, the decline in precision is now much less: for example, the average coefficient of variation for a long position falls by under 30% as the confidence level rises from 90% to 99%. This is consistent with what we would expect, bearing in mind that the coefficient of variation reflects two offsetting factors – an increased point estimate of VaR in the numerator and an increased standard error in the denominator – as the confidence level rises.

A natural alternative to the coefficient of variation is a confidence interval. We can make the confidence interval into an indicator of relative precision if we work with standardised confidence intervals by dividing the bounds of the confidence interval by the point estimate (given in (a)). The confidence interval provides a readily understood indicator of precision and it would highlight possible asymmetry between upside and downside precision in the form of an asymmetric confidence interval. Estimates of standardized 90% confidence intervals are given in section (d) of Table 2. This shows that the bounds of the confidence intervals are within +/- 5% for VaR at the 90% confidence level, within +/- 6 or 7% for VaR at the 95% confidence level, and somewhat wider (and more variable both across contracts and across trading positions) for the 99% confidence level. Thus, the broad picture (although there are exceptions) is that we get modest increases in confidence intervals as the confidence level rises, but there is also considerable variation across contracts and a certain amount of variation across positions.

## 6. RESULTS FOR EXPECTED SHORTFALL

Table 3 presents corresponding non-parametric results for the ES as a risk measure. The point ES estimates are shown in section (a). These are higher than the earlier VaRs, as we would expect: the average ES across all contracts is 3.62%, which compares to an average VaR of 2.57%. However, the ES estimates otherwise show much the same pattern as the VaR estimates: they vary across contract, increase with confidence level and, like the VaR, approximately double from the 90% to 99% confidence levels. In addition, we again find that the Hang Seng is the most risky contract, and the S&P and FTSE are the safest.

**Insert Table 3 here**

Table 3 also presents results for our three precision metrics as they relate to the ES. The standard errors in section (b) are on average about 70% higher than those for the VaR, but otherwise show much the same behavior (they increase markedly with the confidence level, etc.). The coefficients of variation in section (c) are a little lower than the VaR ones, and the 90% confidence intervals in section (d) are on average much the same as the VaR ones. There is some variation across the ES results, and also between the VaR and ES ones, but the patterns of variation are otherwise quite similar. If we use the standard error, we would conclude that the ES estimates are about 70% less precise than the VaR ones, but if we use the other (arguably more reliable) precision metrics, we would conclude that there is little or no difference in the precision of ES estimators compared to VaR ones.[9]

## 7. RESULTS FOR SPECTRAL RISK MEASURES

Table 4 shows non-parametric results for the spectral risk measure. Perhaps the result that stands out most from this Table is that estimates of the spectral risk measure rise quite strongly with the degree of absolute risk aversion (ARA). As a broad order of magnitude, increasing the degree of ARA from 5 to 80 leads estimates of the spectral risk measure to increase by about 300%. At one extreme, the spectral risk estimates with ARA=5 are about the same order of magnitude as the VaR at the 90% confidence level; and, at the other, the spectral risk estimates with ARA=80 are much the same as the ES at the 99% confidence level. In other words, the spectral risk estimates tend to fall in the range encompassed by the estimates of our earlier risk measures. Our results also show that, whilst there is variation across contracts, there is virtually no difference between estimates for long and short positions.

The precision estimates in sections (b)-(d) indicate that precision tends to fall as ARA gets larger. This would make sense as an increasing ARA would suggest that we are placing more and more weight on a smaller number of extreme observations, indicating that the effective sample size is falling. This suggests that the ARA plays much the same role in spectral risk measures as the confidence level does for the VaR and ES. In addition, estimates of spectral risk measures with ARA in the (quite wide) range [5, 80] are of comparable precision to estimates of VaR based on confidence levels in the (quite wide) range [90%, 99%]: crudely put, *estimates of spectral risk measures are of much the same precision as estimates of VaR*.

**Insert Table 4 here**

## 8. CONCLUSIONS

This paper presents some non-parametric estimates of three alternative risk measures applied to various equity futures market positions. The paper compares the VaR, the ES, and spectral risk measures based on an exponential risk-aversion function. Our empirical findings suggest that there is considerable similarity across the three different risk measures. In all three cases, the most important factor determining the magnitude of the risk measures is the conditioning parameter – the confidence level for the VaR and ES, and the coefficient of absolute risk aversion for the spectral risk measures. The orders of magnitude of the risk measures are also quite close – in fact, over the range of parameters considered, we find that the estimated spectral risk measures are somewhere between the lowest estimated VaR and the highest estimated ES. We also estimated various precision indicators for our risk measures, and these also paint a fairly consistent story – that estimates of ES measures are perhaps a little less precise than estimates of VaR ones, and that estimates of spectral risk measures are of much the same precision as VaR estimates. How far these results might apply in other contexts is an open empirical question, but at least the results presented here give us a solid empirical example of how estimators of these measures and of their precision indicators compare with each other in the important case of positions in equity futures contracts.[10]

**NOTES**

1       Some estimates of extreme-value spectral risk measures are presented in Cotter and Dowd (2006), but these estimates must be interpreted in their EV context. In contrast the analysis of empirical properties of VaR and ES is more developed (for example, see Pritsker (1997), Butler and Schachter (1998), Yamai and Yoshiba (2002), Giannopoulos and Tunaru (2004), Chen and Tang (2005), and Gourieroux and Liu (2006)).

2       There is of course a second conditioning parameter, namely the holding or horizon period. However, this plays a passive role in our analysis as we restrict ourselves to a given daily horizon.

3       There are many more applications that rely on quantile based market risk measures such as VaR including price limits and minimum capital requirements. Given the potential risk associated with futures trading it is not surprising that there is an extensive literature looking at these applications and/or the use of these risk measures for futures including (and by no means exhaustive): Hsieh (1993), Broussard (2001), Longin (2001), Cotter, 2004, Werner and Upper (2004) and Brooks et al (2005).

4       Commercial adjustments such as incorporating levels of liquidity could lead to adjustments to any initial margins. SPAN systems associate the margin to cover a large proportion of price movements, for example 99%, on a family of contracts with the same underlying. For further details see Artzner et al., 1999; and London Clearing House, 2002.

5       We note here that the user in a futures market context might be individual investor or the clearinghouse itself. In the former case, the risk measures are those faced by the investors, and in the latter case they are those faced by the clearinghouse. This latter case is where we might use the risk measures to set margin requirements.

6       The vanilla bootstrap used is as explained in the early chapters of Efron and Tibshirani (1993): each resample observation is drawn with equal probability from the sample and then placed back in the sample, and there are no adjustments of confidence bound estimates for possible bias. Such refinements are not necessary in the present context because bias is essentially a small sample problem that does not arise with the large samples available to us here. Estimates reported in the paper are based on bootstrap trials with 5000 sets of resamples.

7       In addition, as noted earlier in the introduction, a non-parametric approach has the nice feature that it allows us to interpret the resulting risk measure estimates as potential estimates of initial margin requirements.

8       Precision metrics are sometimes much more difficult to obtain using parametric approaches. For example, there are few simple expressions for the confidence intervals of any of the risk measures considered here, and the only practical alternative is to rely on the theory of order statistics (e.g., as in Dowd (2001)). However, order-statistics approaches are more difficult to implement than the non-parametric bootstrap used here, and can only be applied to risk measures based on probabilistic conditioning parameters. This means that they can be used to estimate the confidence intervals of VaR or ES, but not the confidence intervals associated with SRMs.

9      This finding is consistent with earlier literature but also goes beyond it. For example, Acerbi (2004, pp. 200-205) finds that the ES typically has a standard error larger but not too much larger than that of the VaR, and our findings are consistent with his. However, he only compared standard errors, and our findings suggest that the other precision metrics (i.e., the coefficient of variation and the confidence intervals) give results that are more favorable to the ES relative to the VaR.

10     There are many possible extensions to this paper. Within the confines of a non-parametric approach, it would be very interesting to extend the analysis to encompass other forms of spectral risk measure based on alternative risk-aversion functions (e.g., power and HARA functions). Going outside the non-parametric paradigm, there are also natural extensions to parametric (e.g., GARCH) and semi-parametric (e.g., filtered historical simulation) modelling. Such approaches are more difficult to handle, but correctly applied, may also be more powerful than the non-parametric methods used here.

**FIGURES**

**Figure 1: Exponential Risk Aversion Functions**

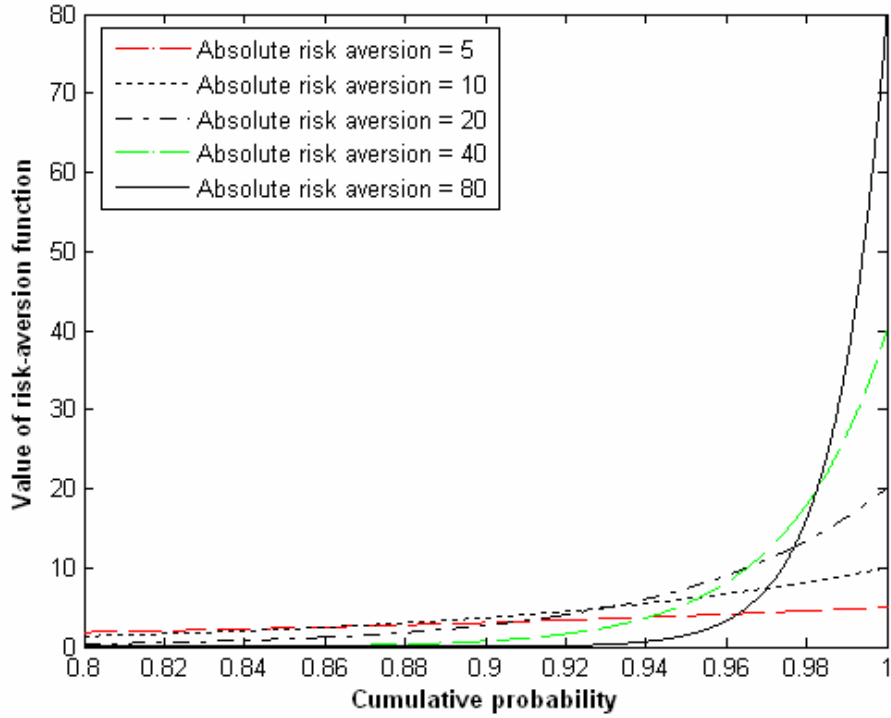

Notes: Based on equation (4) in the text, for stated values of the degree of absolute risk aversion (*k*). This Figure shows how the weights rise with the cumulative probability *p* where $p \in [0.8,1]$, and the rate of increase depends on *k*: the more risk-averse the user, the more rapidly the weights rise.

**TABLES**

| Table One: Summary Statistics for Return Data | | | | | |
|---|---|---|---|---|---|
| | **S&P** | **FTSE** | **DAX** | **HANGSENG** | **NIKKEI** |
| Mean | 0.0004 | 0.0002 | 0.0003 | 0.0004 | -0.0002 |
| Std Dev | 0.0109 | 0.0115 | 0.0150 | 0.0191 | 0.0148 |
| Skewness | -0.1044 | -0.0520 | -0.2616 | 0.3307 | 0.0561 |
| Kurtosis | 7.4198 | 5.3209 | 7.4114 | 13.8301 | 4.7839 |
| $n$ | 3392 | 3392 | 3392 | 3392 | 3392 |
| Minimum | -0.0776 | -0.0606 | -0.1285 | -0.1609 | -0.0760 |
| Maximum | 0.0575 | 0.0595 | 0.0838 | 0.2298 | 0.0800 |

Notes: Based on the 3392 close of day returns for each of the stated indexes over the period January 1 1991 to December 31 2003.

## Table 2: Estimates of VaRs and Precision of VaRs of Futures Positions

| | S&P | FTSE | DAX | HANGSENG | NIKKEI | Mean |
|---|---|---|---|---|---|---|
| **(a) VaR estimates** | | | | | | |
| *Long position* | | | | | | |
| 90% VaR | 0.0119 | 0.0135 | 0.0169 | 0.0208 | 0.0173 | 0.0161 |
| 95% VaR | 0.0175 | 0.0185 | 0.0236 | 0.0294 | 0.0239 | 0.0226 |
| 99% VaR | 0.0319 | 0.0304 | 0.0404 | 0.0512 | 0.0371 | 0.0382 |
| *Short position* | | | | | | |
| 90% VaR | 0.0117 | 0.0131 | 0.0160 | 0.0203 | 0.0177 | 0.0158 |
| 95% VaR | 0.0181 | 0.0175 | 0.0243 | 0.0294 | 0.0247 | 0.0228 |
| 99% VaR | 0.0291 | 0.0329 | 0.0425 | 0.0520 | 0.0368 | 0.0387 |
| | | | | | Overall mean | 0.0257 |
| **(b) Standard errors of VaR** | | | | | | |
| *Long position* | | | | | | |
| 90% VaR | 0.0004 | 0.0004 | 0.0005 | 0.0006 | 0.0005 | 0.0005 |
| 95% VaR | 0.0006 | 0.0006 | 0.0008 | 0.0010 | 0.0008 | 0.0008 |
| 99% VaR | 0.0018 | 0.0015 | 0.0020 | 0.0020 | 0.0012 | 0.0017 |
| *Short position* | | | | | | |
| 90% VaR | 0.0004 | 0.0003 | 0.0006 | 0.0005 | 0.0005 | 0.0004 |
| 95% VaR | 0.0006 | 0.0005 | 0.0008 | 0.0008 | 0.0006 | 0.0007 |
| 99% VaR | 0.0013 | 0.0014 | 0.0022 | 0.0026 | 0.0015 | 0.0018 |
| | | | | | Overall mean | 0.0010 |
| **(c) Coefficients of variation of VaR** | | | | | | |
| *Long position* | | | | | | |
| 90% VaR | 26.86 | 32.47 | 37.29 | 25.60 | 35.68 | 31.58 |
| 95% VaR | 29.40 | 31.75 | 30.70 | 29.40 | 28.80 | 30.01 |
| 99% VaR | 17.72 | 20.27 | 20.20 | 25.60 | 30.92 | 22.94 |
| *Short position* | | | | | | |
| 90% VaR | 32.48 | 43.07 | 29.03 | 40.33 | 37.12 | 36.41 |
| 95% VaR | 29.57 | 34.17 | 29.70 | 35.87 | 40.35 | 33.93 |
| 99% VaR | 22.38 | 23.50 | 19.32 | 20.00 | 24.53 | 21.95 |
| | | | | | Overall mean | 29.47 |
| **(d) 90% confidence intervals for VaR** | | | | | | |
| *Long position* | | | | | | |
| 90% VaR | [0.9518  1.0649] | [0.9547  1.0492] | [0.9594  1.0467] | [0.9542  1.0483] | [0.9501  1.0444] | |
| 95% VaR | [0.9381  1.0508] | [0.9493  1.0533] | [0.9521  1.0551] | [0.9474  1.0561] | [0.9556  1.0642] | |
| 99% VaR | [0.8792  1.0971] | [0.9196  1.0783] | [0.9064  1.0844] | [0.9309  1.0672] | [0.9619  1.0872] | |
| *Short position* | | | | | | |
| 90% VaR | [0.9422  1.0479] | [0.9660  1.0437] | [0.9556  1.0633] | [0.9590  1.0356] | [0.9524  1.0458] | |
| 95% VaR | [0.9343  1.0482] | [0.9465  1.0420] | [0.9496  1.0528] | [0.9544  1.0447] | [0.9579  1.0487] | |
| 99% VaR | [0.9298  1.0731] | [0.8980  1.0579] | [0.9242  1.0927] | [0.9113  1.0693] | [0.9342  1.0578] | |

*Notes*: VaR estimates in daily return terms are based on the average of 5000 bootstrap resamples. The precision estimates, standard error of VaR, coefficients of variation of VaR and standardised 90% confidence intervals of VaR are also based on 5000 bootstrap resamples. The holding period is 1 day. Bounds of confidence intervals are standardised (i.e., divided) by the means of the estimates.

| | S&P | FTSE | DAX | HANGSENG | NIKKEI | Mean |
|---|---|---|---|---|---|---|
| **Table 3: Estimates of ESs and Precision of ESs of Futures Positions** | | | | | | |
| **(a) ES estimates** | | | | | | |
| *Long position* | | | | | | |
| 90% ES | 0.0200 | 0.0209 | 0.0269 | 0.0345 | 0.0270 | 0.0259 |
| 95% ES | 0.0256 | 0.0261 | 0.0340 | 0.0445 | 0.0336 | 0.0328 |
| 99% ES | 0.0408 | 0.0387 | 0.0527 | 0.0699 | 0.0472 | 0.0499 |
| *Short position* | | | | | | |
| 90% ES | 0.0200 | 0.0205 | 0.0278 | 0.0341 | 0.0270 | 0.0259 |
| 95% ES | 0.0254 | 0.0260 | 0.0358 | 0.0440 | 0.0329 | 0.0328 |
| 99% ES | 0.0385 | 0.0398 | 0.0544 | 0.0699 | 0.0476 | 0.0500 |
| | | | | | Overall mean | 0.0362 |
| **(b) Standard errors of ES** | | | | | | |
| *Long position* | | | | | | |
| 90% ES | 0.00061 | 0.00055 | 0.00077 | 0.00120 | 0.00072 | 0.0008 |
| 95% ES | 0.00090 | 0.00081 | 0.00120 | 0.00200 | 0.00098 | 0.0012 |
| 99% ES | 0.00200 | 0.00180 | 0.00280 | 0.00650 | 0.00240 | 0.0031 |
| *Short position* | | | | | | |
| 90% ES | 0.00063 | 0.00058 | 0.00089 | 0.00110 | 0.00068 | 0.0008 |
| 95% ES | 0.00089 | 0.00092 | 0.00130 | 0.00170 | 0.00091 | 0.0011 |
| 99% ES | 0.00260 | 0.00170 | 0.00340 | 0.00470 | 0.00240 | 0.0030 |
| | | | | | Overall mean | 0.0017 |
| **(c) Coefficients of variation of ES** | | | | | | |
| *Long position* | | | | | | |
| 90% ES | 32.53 | 37.80 | 34.98 | 28.75 | 37.61 | 34.34 |
| 95% ES | 28.46 | 32.28 | 28.33 | 22.25 | 34.29 | 29.12 |
| 99% ES | 20.40 | 21.50 | 18.82 | 10.75 | 19.67 | 18.23 |
| *Short position* | | | | | | |
| 90% ES | 31.61 | 35.44 | 31.09 | 31.00 | 39.68 | 33.76 |
| 95% ES | 28.60 | 28.38 | 27.54 | 25.88 | 36.14 | 29.31 |
| 99% ES | 14.81 | 23.41 | 16.00 | 14.87 | 19.83 | 17.79 |
| | | | | | Overall mean | 27.09 |
| **(d) 90% confidence intervals for ES** | | | | | | |
| *Long position* | | | | | | |
| 90% ES | [0.9496 1.0513] | [0.9569 1.0431] | [0.9538 1.0478] | [0.9472 1.0567] | [0.9562 1.0436] | |
| 95% ES | [0.9435 1.0596] | [0.9486 1.0517] | [0.9449 1.0577] | [0.9308 1.0759] | [0.9511 1.0485] | |
| 99% ES | [0.9206 1.0806] | [0.9239 1.0768] | [0.9121 1.0919] | [0.8632 1.1668] | [0.9217 1.0884] | |
| *Short position* | | | | | | |
| 90% ES | [0.9470 1.0528] | [0.9545 1.0472] | [0.9482 1.0541] | [0.9473 1.0546] | [0.9592 1.0420] | |
| 95% ES | [0.9439 1.0586] | [0.9427 1.0588] | [0.9417 1.0611] | [0.9387 1.0650] | [0.9548 1.0456] | |
| 99% ES | [0.8937 1.1193] | [0.9326 1.0729] | [0.9042 1.1069] | [0.8936 1.1160] | [0.9188 1.0852] | |

*Notes*: ES estimates in daily return terms are based on an average from 5000 bootstrap resamples. The precision estimates, standard error of ES, coefficients of variation of ES and standardised 90% confidence intervals of ES are also based on 5000 bootstrap resamples. The holding period is 1 day. Bounds of confidence intervals are standardised (i.e., divided) by the means of the estimates.

**Table 4: Non-parametric Estimates of Spectral Risk Measures and Precision of Spectral Risk Measures of Futures Positions**

|  | S&P | FTSE | DAX | HANGSENG | NIKKEI | Mean |
|---|---|---|---|---|---|---|
| **(a) Spectral risk measure estimates** | | | | | | |
| *Long position* | | | | | | |
| ARA= 5 | 0.0116 | 0.0124 | 0.0158 | 0.0200 | 0.0156 | 0.0151 |
| ARA = 10 | 0.0170 | 0.0178 | 0.0228 | 0.0293 | 0.0227 | 0.0219 |
| ARA = 20 | 0.0225 | 0.0231 | 0.0300 | 0.0390 | 0.0295 | 0.0288 |
| ARA = 40 | 0.0284 | 0.0284 | 0.0374 | 0.0495 | 0.0360 | 0.0359 |
| ARA = 80 | 0.0349 | 0.034 | 0.0454 | 0.0613 | 0.0424 | 0.0436 |
| *Short position* | | | | | | |
| ARA= 5 | 0.0111 | 0.0120 | 0.0156 | 0.0192 | 0.0159 | 0.0148 |
| ARA = 10 | 0.0166 | 0.0174 | 0.0232 | 0.0286 | 0.0228 | 0.0217 |
| ARA = 20 | 0.0222 | 0.0229 | 0.0311 | 0.0384 | 0.0293 | 0.0288 |
| ARA = 40 | 0.0278 | 0.0286 | 0.0392 | 0.0489 | 0.0355 | 0.0360 |
| ARA = 80 | 0.0338 | 0.0347 | 0.0477 | 0.0603 | 0.0420 | 0.0437 |
| | | | | | Overall mean | 0.0290 |
| **(b) Standard errors of spectral risk measure** | | | | | | |
| *Long position* | | | | | | |
| ARA= 5 | 0.0003 | 0.0003 | 0.0004 | 0.0006 | 0.0004 | 0.0004 |
| ARA = 10 | 0.0005 | 0.0005 | 0.0006 | 0.0010 | 0.0006 | 0.0006 |
| ARA = 20 | 0.0008 | 0.0007 | 0.0010 | 0.0017 | 0.0008 | 0.0010 |
| ARA = 40 | 0.0011 | 0.0009 | 0.0014 | 0.0028 | 0.0012 | 0.0015 |
| ARA = 80 | 0.0015 | 0.0013 | 0.0019 | 0.0049 | 0.0017 | 0.0023 |
| *Short position* | | | | | | |
| ARA= 5 | 0.0003 | 0.0003 | 0.0005 | 0.0006 | 0.0004 | 0.0004 |
| ARA = 10 | 0.0005 | 0.0005 | 0.0007 | 0.0009 | 0.0006 | 0.0006 |
| ARA = 20 | 0.0008 | 0.0007 | 0.0011 | 0.0014 | 0.0008 | 0.0010 |
| ARA = 40 | 0.0012 | 0.0010 | 0.0017 | 0.0022 | 0.0011 | 0.0014 |
| ARA = 80 | 0.0019 | 0.0014 | 0.0026 | 0.0034 | 0.0016 | 0.0022 |
| | | | | | Overall mean | 0.0011 |
| **(c) Coefficients of variation of spectral risk measure** | | | | | | |
| *Long position* | | | | | | |
| ARA= 5 | 34.85 | 38.57 | 35.72 | 31.29 | 37.69 | 35.63 |
| ARA = 10 | 33.08 | 38.11 | 35.34 | 29.30 | 38.53 | 34.87 |
| ARA = 20 | 29.44 | 33.95 | 31.49 | 22.94 | 36.41 | 30.85 |
| ARA = 40 | 25.82 | 29.96 | 26.71 | 17.68 | 30.00 | 26.03 |
| ARA = 80 | 23.27 | 26.15 | 23.89 | 12.51 | 24.94 | 22.15 |
| *Short position* | | | | | | |
| ARA= 5 | 32.37 | 36.09 | 31.89 | 31.19 | 39.54 | 34.22 |
| ARA = 10 | 32.18 | 35.59 | 31.74 | 30.97 | 40.69 | 34.24 |
| ARA = 20 | 28.30 | 32.02 | 28.27 | 27.43 | 37.10 | 30.63 |
| ARA = 40 | 23.17 | 28.60 | 23.06 | 22.23 | 32.27 | 25.87 |
| ARA = 80 | 17.79 | 24.79 | 18.35 | 17.74 | 26.25 | 20.98 |
| | | | | | Overall mean | 29.55 |

**(d) 90% confidence intervals for spectral risk measure**

|  | | *Long position* | | | |
|---|---|---|---|---|---|
| ARA= 5 | [0.9534 1.0485] | [0.9564 1.0423] | [0.9541 1.0458] | [0.9483 1.0544] | [0.9562 1.0446] |
| ARA = 10 | [0.9510 1.0512] | [0.9587 1.0444] | [0.9533 1.0452] | [0.9458 1.0594] | [0.9572 1.0416] |
| ARA = 20 | [0.9450 1.0565] | [0.9520 1.0486] | [0.9492 1.0522] | [0.9334 1.0734] | [0.9551 1.0456] |
| ARA = 40 | [0.9359 1.0653] | [0.9445 1.0553] | [0.9391 1.0615] | [0.9146 1.0983] | [0.9484 1.0535] |
| ARA = 80 | [0.9285 1.0708] | [0.9382 1.0650] | [0.9298 1.0716] | [0.8807 1.1420] | [0.9367 1.0661] |
|  | | *Short position* | | | |
| ARA= 5 | [0.9500 1.0516] | [0.9544 1.0456] | [0.9489 1.0521] | [0.9483 1.0535] | [0.9588 1.0420] |
| ARA = 10 | [0.9498 1.0516] | [0.9540 1.0465] | [0.9487 1.0513] | [0.9465 1.0535] | [0.9600 1.0403] |
| ARA = 20 | [0.9424 1.0591] | [0.9496 1.0517] | [0.9434 1.0601] | [0.9405 1.0609] | [0.9563 1.0447] |
| ARA = 40 | [0.9323 1.0740] | [0.9401 1.0607] | [0.9331 1.0737] | [0.9274 1.0775] | [0.9480 1.0524] |
| ARA = 80 | [0.9107 1.0978] | [0.9342 1.0645] | [0.9174 1.0949] | [0.9090 1.0936] | [0.9373 1.0635] |

*Notes*: Risk estimates in daily return terms are based on an average of 5000 bootstrap resamples. The precision estimates, standard error, coefficients of variation and standardised 90% confidence intervals are also based on 5000 bootstrap resamples. The holding period is 1 day. Bounds of confidence intervals are standardised (i.e., divided) by the means of the estimates.